\documentclass[11pt]{article}
\usepackage{sao1}
\usepackage{psfig}

\advance\textheight by\topskip 

\def\degr{\hbox{$^\circ$}}

\def\arcsec{\hbox{$^{\prime\prime}$}}
\def\fm{\hbox{$.\!\!^{\rm m}$}}

\begin{document}

\title{RC J0311+0507: A Candidate for Superpowerful Radio Galaxies
in the Early Universe at Redshift $z$=4.514
\footnote{
ISSN 1063-7737, \\
Astronomy Letters, 2006, Vol.32, No.7, pp.433-438\\
Pleiades Publishing, Inc., 2006. Original Russian Text\\
{\it A.I. Kopylov, W.M. Goss, Yu.N. Pariiskij, N.S. Soboleva,
O.V. Verkhodanov, A.V. Temirova, O.P. Zhelenkova,} 2006,
published in Pisma v Astronomicheskii Zhurnal, 2006, Vol.{\bf 32},
No.7, pp. 483-488.
}
}

\author{
A. I. Kopylov\footnote{e-mail: akop@sao.ru} \inst{a}\and
W. M. Goss,                                 \inst{b}\and
Yu. N. Pariiskii,                           \inst{a}\and
N. S. Soboleva,                             \inst{c}\and
O. V. Verkhodanov,                          \inst{a}\and
A. V. Temirova,                             \inst{c}\and
O. P. Zhelenkova                            \inst{a}
}

\institute{
\saoname,
\and
National Radio Astronomy Observatory, 520 Edgemont Road, Charlottesville,
     VA 22903, USA,
\and
St. Petersburg Branch of the Special Astrophysical Observatory,
     Russian Academy of Sciences, Pulkovo,
    St. Petersburg, 196140 Russia
}
\date{August 31, 2005}{}
\maketitle{}

\begin{abstract}
A strong emission line at 6703\AA has been detected in the optical 
spectrum for the host galaxy
(R=23.1) of the radio source RC\,J0311+0507 (4C+04.11). This radio
galaxy, with a spectral index of 1.31
in the frequency range 365--4850 MHz, is one of the ultrasteep spectrum
objects from the deep survey of
a sky strip conducted with RATAN-600 in 1980--1981. We present
arguments in favor of the identification
of this line with Ly$\alpha$ at redshift $z=4.514$. In this case, the object
belongs to the group of extremely distant
radio galaxies of ultrahigh radio luminosity
($P_{1400}=1.3\times10^{29}W Hz^{-1}$).
Such power can be provided
only by a fairly massive black hole ($\sim10^9M_{\sun}$)
that formed in a time
less than the age of the Universe at the 
observed $z$ (1.3 Gyr) or had a primordial origin.


\end{abstract}

{\bf DOI}: 10.1134/S1063773706070012

\section{Introduction}
The radio source RC J0311+0507 (the RATAN
Cold Catalog; Parijskij et al. 1991, 1992) was dis
covered in 1980--1981 observations during the first
deep survey of a sky strip with the RATAN-600 multifrequency complex
(Berlin et al. 1981). The catalog included more than 1145 radio sources with
a flux density limit higher than 10 mJy at 7.6 cm.
The RATAN-600 observations at various azimuths
allowed a positional accuracy of $\sim$15\arcsec to be obtained.
This accuracy is not enough for deep optical identi-
fications, but is quite sufficient for deep VLA
observations. The absence of catalogs with an adequate
sensitivity in those years made it difficult to identify
them with known objects. The first catalog of a high
positional accuracy with a sensitivity up to 200 mJy
was the UTRAO (Texas) Catalog at $\sim$80 cm (Douglas et al. 1996).
Douglas kindly provided us data on
our surveyed area long before the publication of this
catalog. This allowed us to identify at least the objects
of the RC catalog with fairly steep spectra. There were
about one-third of these sources. RC J0311+0507
was one of them. Since its spectral index ($\sim\nu^{-\alpha}$)
is $\alpha\approx1.2$, it was included in the subsample of candidates for
distant objects of the Big Trio Project (Goss et al. 1994;
Kopylov et al. 1995; Parijskij et al. 1999;
Verkhodanov et al. 2001).

Note that RC J0311+0507 is a fairly bright low-frequency radio source.
It was first detected at a frequency of 85 MHz (Mills et al. 1958) and was
then reliably recorded at 178 MHz (Gower et al. 1967) as the
object 4C+04.11 with a flux density of 5.5\,Jy. R\"ottgering et
al. (1994) independently selected RC\,J0311+0507
to be included in their sample of objects with steep
radio spectra (365B B0309+049).
However, subsequently they did not study it in the
optical range, possibly because of an uncertain spectral index.
RC\,J0311+0507 also closely corresponds
in its parameters to the objects of the sample of
steep-spectrum radio sources by Tielens et al. (1979).

Analysis of this sample revealed the then most distant
radio galaxy, 4C+41.17 (z=3.80, Chambers et al. 1990).

\begin{figure}[!th]
\centerline{\psfig{figure=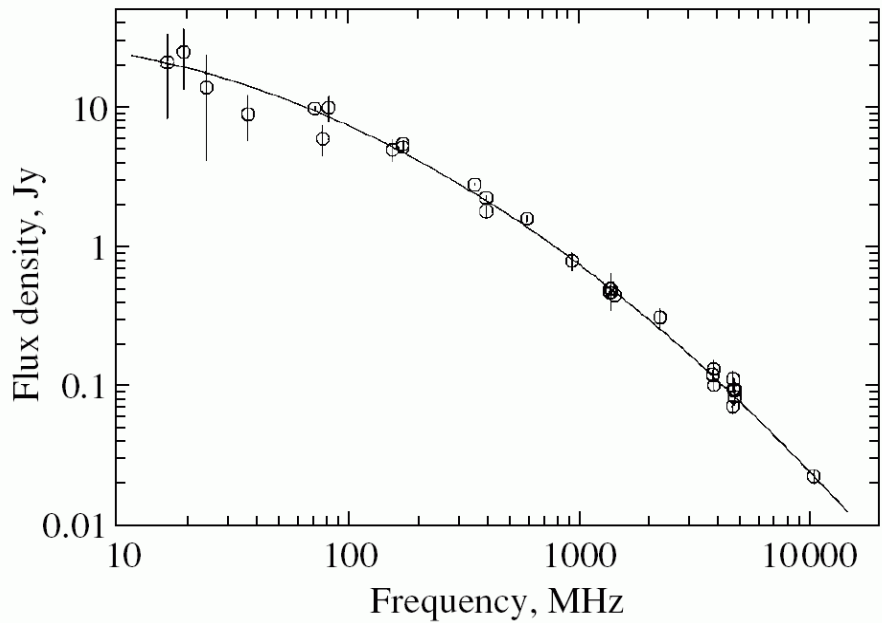,width=12cm,angle=-90}}
\caption{
Radio spectrum of RC J0311+0507 constructed
from the data accumulated by 2005. The spectral index in 
the frequency range 365-4850 MHz is 1.31. The spectrum flattening
toward the low frequencies suggests that
the components of the radio source are compact. 
}
\end{figure}

We have studied the object on VLA with a resolution of 1''.4 at 21\,cm as
part of the Big Trio Project
(RATAN--VLA--BTA). The radio source turned out
to be compact, about 2'', with an AD (Asymmetric
Double) structure. The VLA archival data with a 
resolution of 0''.4 at 6\,cm show the presence of a
third, very weak component of small angular size. 
Below, we provide the main data on this object, including the radio 
data and optical studies with the 
6-m BTA telescope of the Special Astrophysical Observatory (SAO) 
(identification, multicolor photometry, and spectroscopy). 

\section{Observations}

\subsection{Radio Observations}

Figure 1 shows the radio spectrum of 
RC J0311+0507 with all of the available measurements collected in the 
CATS database (Verkhodanov et al. 1997), including the RATAN-600
multifrequency data. We also
added the measurements at 38 
and 178\,MHz from Williams et al. (1968). The curve
in Fig.\,1 corresponds to the equation
$$
\log S=1.423+0.212\log\nu-0.241(\log\nu)^2            (1)
$$
that was obtained by fitting a parabola to all measurements (31 data 
points). In the frequency range 
365--4850\,MHz, the object has an ultrasteep spectrum ($\alpha$=1.31), which
is the first signature of a high
redshift. The increase in the spectral slope from low 
to high frequencies is also a characteristic property of 
distant compact powerful radio sources. 

\begin{figure}[!th]
\centerline{\psfig{figure=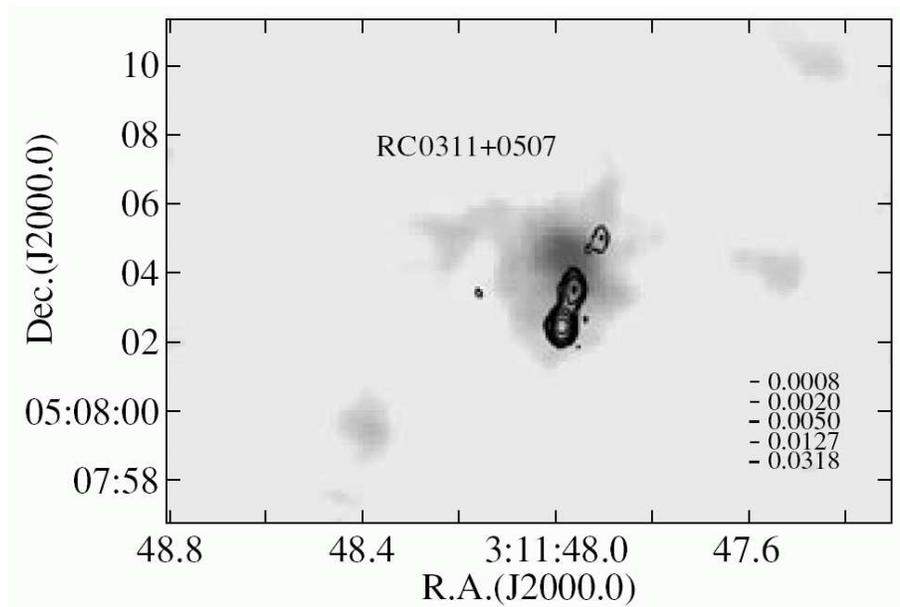,width=12cm,angle=-90}}
\caption{Superposition of the 4860\,MHz VLA isophotal
image of the radio source RC\,J0311+0507 obtained in
1985 on the R-band BTA 1995 image of the host galaxy. 
}
\end{figure}

The VLA observations carried out by W.M. Goss 
in June 1995 with a resolution of 1''.4 at 1425\,MHz
provided evidence for a compact two-component 
structure (Parijskij et al. 1996). Based on their VLA 
observations with a similar resolution, Rottgering 
et al. (1994) determined the radio source as an 
extended one with a size of 1''.6. With the kind
permission of B.\,Burke, we found the image of this
object obtained by J.\,Hewitt in 1985 at 4860\,MHz
with a resolution of 0''.4 in the VLA archive where
a linear triplet structure with a total angular size 
of 2''.8 is seen. Based on these data, we can classify RC\,J0311+0507 as
a compact steep-spectrum
(CSS) object. It is distinguished by a large flux 
density asymmetry ($\sim$20 times) between the two
extreme components, which is much more commonly 
observed in quasars than in radio galaxies. 

\subsection{Optical Identification}

Based on the direct BTA image with an exposure 
time of 400\,s at 2'' seeing obtained in September
1995 with a 580$\times$520 ISD015 CCD array (pixel
size 0''.205$\times$0''.154), we identified the radio source
(Parijskij et al. 1996) with a faint galaxy (R$\approx$22.9 in
a 5'' aperture). The optical-to-radio luminosity ratio
turned out to be the standard one for the population 
of luminous radio galaxies (McCarthy 1993; Parijskij 
et al. 1996). Figure 2 shows the optical object with 
the superimposed VLA 4860\,MHz isophotes.

\subsection{Multicolor Photometry}

In November 1999, we performed the next observations with the PMCCD 
instrument (a TX1024A 
array with 0''.206$\times$0''.206 pixels).
We obtained B, V, R, and I images with exposure times of 600, 1000,
400, and 1000\,s, respectively, at $\sim$2'' seeing.
The photometric measurements with a 5'' aperture corrected
for the extinction in the Galaxy
($A_B$=0.83, $A_V$=0.64, $A_R$=0.51, and $A_I$=0.37) yielded magnitudes
of $>$24.9, 24.8$\pm$0.6, 22.6$\pm$0.15,
and 22.3$\pm$0.4, respectively. The
color characteristics are close to those
expected for massive galaxies at $z=3-5$, and the
complete absence of the object in the B band does not
contradict to the emission cutoff beyond the Lyman 
912\,\AA limit. The R-band size of the galaxy slightly
exceeds its I-band size. This may suggest the presence of a hydrogen 
halo around the host galaxy that 
is commonly observed in distant radio galaxies. The 
presence of a halo can lead to a considerable increase 
in the object's size if the strong Ly$\alpha$ line falls not far
from the passband maximum of the corresponding filter
(see, e.g., RC\,J0105+0501; Soboleva et al. 2000),
where the Ly$\alpha$ line increases significantly the V-band
size of the object). 

\subsection{Spectroscopic Observations}

In September and November 2004, we obtained 
BTA spectra of the host galaxy. The observations were 
carried out with the SCORPIO universal focal reducer that was put into 
operation on BTA late in 2003 
as the main multipurpose, high-efficiency instrument 
(Afanasiev and Moiseev 2005). On November 8--9,
2004, we were able to obtain the best-quality spectrum of the host 
galaxy with a total exposure time of 
3600\,s in long-slit observations at 1'' seeing. The
gr300G grating provided the entire spectral range 
accessible to the instrument (3800--9400\,\AA) with a
resolution of $\sim$20\,\AA, which is commonly used to study
objects of this type. The slit width was 1'' and the
position angle was $-$11\degr. The spectrum was reduced
using the SCORPIO data reduction and analysis 
software package (Afanasiev and Moiseev 2005) and 
is shown in Fig.\,3. The size of the region of integration
over the slit height was 1''.6. The absolute spectrum
calibration was performed using the spectrophotometric standard Hiltner 
600 and is given in units of 
10$^{-17}$ erg cm$^{-2}$ s$^{-1}$\AA$^{-1}$.

An intense line is seen at a wavelength of 6703\,\AA.
The line flux is $\approx$5$\times10^{-16}$erg cm$^{-2}$s$^{-1}$,
the $FWHM$
is $\sim$1500\,km s$^{-1}$, and the equivalent width is
$\sim$1000\,\AA. We interpret it as Ly$\alpha$ with a redshift of
4.514$\pm$0.001. The luminosity in this line is close
to the (R-band) continuum luminosity, which is 
observed in steep-spectrum radio galaxies only for 
the Ly$\alpha$ line (McCarthy 1993).

The alternative interpretation ([O II] 3727\,\AA with
$z=0.8$) is highly unlikely because of the complete
absence of [O III] 5007\,\AA,
which is usually twice as
intense as [O II] 3727\,\AA
for this population of objects.
The identification with Ly$\alpha$ is consistent with the
weakness of other lines falling into the spectral range 
studied of which only the C IV 1549\,\AA
line is detected with a signal-to-noise ratio of $\sim$2 at
10\% of the Ly$\alpha$  intensity.

The ratio of the continuum levels in 400\,\AA--wide
intervals on both sides of Ly$\alpha$ is $\sim$3. The lowering
of the continuum at wavelengths shortward of Ly$\alpha$
is attributable to the absorption by the Ly$\alpha$ forest
and is in agreement with the data for quasars at 
$z=4.5$(Songaila 2004). In general, the spectrum
of RC\,J0311+0507 is similar in its characteristics
to the spectra of high-redshift radio galaxies (see, 
e.g., 8C\,1435+63, $z=4.261$, Fig.\,1 in Spinrad et al.
(1995)). 

\section{Discussion}

Although $z=4.514$ is considerably lower than the
limiting redshifts detected to date, for galaxies (Malhotra and Rhoads 
2005; Stanway et al. 2003; Pello 
et al. 2004) and quasars (Fan et al. 2003; Walter et al. 
2004), RC\,J0311+0507 is only the second luminous
radio galaxy detected at a redshift higher than 4.5. 

Let us compare the main parameters of 
RC\,J0311+0507 with those of other radio galaxies
at $z>4$. Only seven such galaxies are known and
almost all of them have been studied more or less 
adequately. Table 1 successively lists the names of 
the radio galaxies, their redshifts, optical R (or I) 
magnitudes, infrared K magnitudes, 1400-MHz flux 
densities (NVSS; Condon et al. 1998) (except the object
VLA\,J123642+621331, for which the data were
taken from Richards (2000)), two-frequency spectral 
indices $\alpha$ from the Texas Survey (365\,MHz) and
NVSS (with the exception of VLA J123642+621331, 
for which only 1.4 and 8.5\,GHz measurements are
available; Richards 2000), the largest angular sizes 
(LAS) in arcseconds, and morphology of the radio 
galaxies in the standard notation
(S--single, D--double, AD--asymmetric double,
C--core, and E--extended). The last column gives references to the
publications from which the redshifts, optical magnitudes, infrared 
magnitudes, and LAS of the radio 
sources were taken. 

In three cases (VLA J123642+621331, 
TN J1123--2154, and 7C\,1814+670), only a weak
Ly$\alpha$ line was detected; in the remaining cases, the
Ly$\alpha$ line is very intense. The color data (after the subtraction of the
Ly$\alpha$ contribution, $\sim$0\fm7) are consistent
with new models for evolution of large galaxies. 

\begin{figure}[!th]
\centerline{\psfig{figure=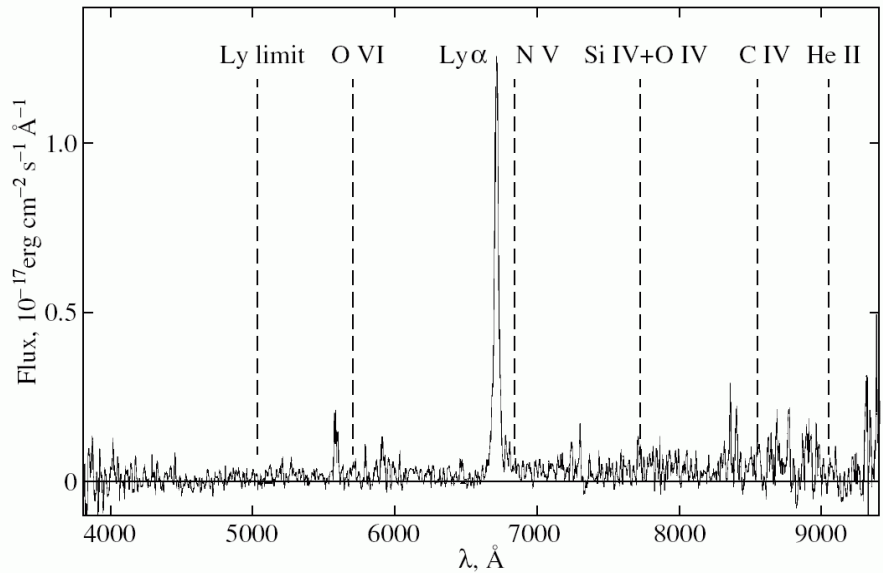,width=12cm,angle=-90}}
\caption{Optical spectrum of the host galaxy of the radio source RC
J0311+0507. We identify a narrow, intense line at the center.
with Ly$\alpha$\,1216\,\AA.
The dashed indicate the expected positions of the
emission lines typical of distant radio galaxies. Other
spectral features include the residual effect of strong atmospheric 
lines after the subtraction of the night-sky spectrum. 
}
\end{figure}

Thus, for example, for the GALEV2 model (Bicker 
et al. 2004) with the assumed epoch of primary star 
formation ($z=5$), the expected colors of the stellar
population of an elliptical galaxy are given in Table\,2.

This table once again confirms that the case with a 
high redshift is correct. In its redshift, 
RC J0311+0507 is second only to the object 
TN J0924--2201 ($z=5.199$), but this source exceeds
it in radio luminosity. Having determined the flux 
density 3.5 Jy at 254 MHz (which corresponds to 
the emission frequency 1400 MHz at $z=4.514$)
using interpolation over the spectrum, we obtain the 
power of the radio source, 1.3$\times10^{29}$W Hz$^{-1}$ (for
$H_0$=70\,km s$^{-1}$Mpc$^{-1}$, $\Omega_M$=0.3, and $\Omega_\Lambda$=0.7).
RC\,J0311+0507 and 8C\,1435+63 turn out to be
similar in their parameters to superpowerful radio 
galaxies at $z>4$, which exceed in luminosity Cyg\,A,
the most powerful nearby radio galaxy, by a factor 
of $\sim$10.

An ultrahigh radio luminosity is a signature of a 
supermassive black hole ($M_{bh}\sim10^{9}M_{\sun}$) at
the center of the host galaxy.
However, in the standard model, 
the time scale of its growth from the time of secondary ionization from 
$\sim10M_{\sun}$ to $M_{bh}\sim10^{9}M_{\sun}$ is only
$\sim$0.5 Gyr. Therefore, the object can be of interest
in connection with the problem of age crisis (Cunha 
and Santos 2004; Loeb and Barcana 2001). Not all 
of the models for the formation of supermassive black 
holes admit such a fast growth of their masses from 
several solar masses to $10^{9}M_{\sun}$; either severe con
straints on the rate of their growth are needed or 
one must accept the version of primordial (pregalactic) large black 
holes forming stellar systems around 
themselves that have been often discussed in recent 
years. 

The weakness of other lines in the spectrum 
(Fig.\,3) suggests that the main emission in the Ly$\alpha$
line is attributable to a halo that is poorly enriched 
with He\,II (1640\,\AA) and C\,IV (1549\,\AA). The upper
limit on He\,II and C\,IV obtained by Dawson et al.
(2004) for distant Ly$\alpha$-emitting galaxies is 15\% (our
value is 10\%). Therefore, we believe that their conclusion about the
primordial nature of the hydrogen gas 
halo not enriched with nuclear reactions in stars of the 
host galaxy or population III stars ($z>10-30$) (Loeb
and Barcana 2001) for this object is also applicable to 
RC\,J0311+0507.

\section{Conclusions}

The goal of this paper is to draw the attention of 
astronomers that are interested in objects of the early 
Universe to the source RC\,J0311+0507.

With expanding capabilities of the BTA optical 
facilities and with refinement of the selection criteria 
by taking into account the international experience, 
increasingly distant objects from the sample of steep-spectrum
RC objects can be studied.

\begin{table}[!h]
\caption{Data for radio galaxies at $z>4$}
\begin{tabular}{llrccclcc}
\hline
Name              & $z$   & $m_{opt}$ &mK&$S_{1400}$,mJy& $\alpha$& LAS  & Morphology &References$^*$: \\
\hline
TN J0924-2201     & 5.199 &  $>$24 R   & 21.7 & 71     &   1.65  & 1''.2 & D          &  1,2,2,3       \\
RC J0311+0507     & 4.514 &  23.1 R    & ...  & 500    &   1.29  & 2.8   & AD+C       &  4,4,--,4      \\
VLA J123642+621331& 4.424 &  24.9 I    & 21.4 & 0.5    &   0.94  & 0.4   & C+E        &  5,5,5,6       \\
6C 0140+326       & 4.413 &  24 I      & 20.0 & 91     &   1.17  & 2.6   & D          &  7,8,9,8       \\
8C 1435+63        & 4.261 &  23.6 I    & 19.5 & 497    &   1.37  & 3.9   & D+C        &  10,10,9,11    \\
TN J1338-1941     & 4.11  &  22.4 R    & 20.0 & 121    &   1.33  & 5.5   & AD+C       &  7,12,12,13    \\
TN J1123-2154     & 4.109 &  $>$24.5 R & 20.3 & 49     &   1.57  & 0.8   & S          &  7,2,2,3       \\
7C 1814+670       & 4.05  &  24.1 R    & 19.4 & 236    &   1.08  & 18.   & D          &  14,14,15,14   \\
\hline
\end{tabular}
\end{table}
References: 1--Venemans et al. (2004), 2--De Breuck et al. (2002),
3--De Breuck et al. (2000), 4--this work, 5--Waddington
et al. (1999), 6--Muxlow et al. (2005), 7--De Breuck et al. (2001),
8--Rawlings et al. (1996), 9--van Breugel et al. (1998),
10--Spinrad et al. (1995), 11--Lacy et al. (1994),
12--De Breuck et al. (2004), 13--De Breuck et al. (1999),
14--Lacy et al. (1999),
15--Lacy et al. (2000).
RATAN--600 surveys deeper than previous ones
have been used to prepare lists of a weaker population 
of ultrasteep-spectrum radio sources and we hope to 
advance further along the redshift scale.

\begin{table}[!h]
\begin{center}
\caption{Observations and the GALEV2 model}
\begin{tabular}{cccc}
\hline
Band  & RC J0311+0507  &   Model    &    Model       \\
      &                &  $z=4.5$   &   $z=0.8$     \\
\hline
B     &   $>$24.9      &  $>$28.84  &   23.86       \\
V     & 24.8$\pm$0.6   &   24.30    &   22.64       \\
R     & 23.3$\pm$0.3   &   23.19    &   21.71       \\
I     & 22.3$\pm$0.4   &   22.23    &   20.28       \\
K     &    ...         &   20.87    &   17.30       \\
\hline
\end{tabular}
\end{center}
\end{table}

\section{Acknowledgements}
We are grateful to A.V. Moiseev, who provided the 
BTA spectroscopic observations with the SCORPIO 
universal focal reducer. This work was supported in 
part by the Russian Foundation for Basic Research 
(project No. 05-02-17521 and 05-07-90139) and a
grant from the Presidium of the St. Petersburg Science Center.
In this study, we used the NASA/IPAC
Extragalactic Database (NED), which is operated 
by the Jet Propulsion Laboratory, California Institute of Technology, 
under contract with the National 
Aeronautics and Space Administration. The National 
Radio Astronomy Observatory is a facility of the National Science 
Foundation operated under cooperative agreement by Associated 
Universities, Inc.

   Translated by G.Rudnitski


\begin{thebibliography}{}

\bibitem{}
V. L. Afanasiev and A. V. Moiseev, Pisma Astron. Zh.
	 31, 214 (2005) [Astron. Lett. 31, 194 (2005)].
\bibitem{}
A. B. Berlin, E. V. Burlaenko, V. Ya. Golnev, et al.,
       Pisma Astron. Zh. 7, 290 (1981).
\bibitem{}
 J. Bicker, U. Fritze, V. Alvensleben, et al., Astron.
	Astrophys. 413, 37 (2004).
\bibitem{}
  K. C. Chambers, G. K. Miley, and W. J. M. van
	   Breugel, Astrophys. J. 363, 21 (1990).
\bibitem{}
   J. J. Condon, W. D. Cotton, E. W. Greisen, et al.,
	  Astron. J. 115, 1693 (1998).
\bibitem{}
J. V. Cunha and R. Santos, Int. J. Mod. Phys. D 13,
	 1321 (2004).
\bibitem{}
 S. Dawson, J. E. Rhoads, S. Malhotra, et al., Astrophys. J. 617, 707
      (2004).
\bibitem{}
  C. De Breuck, W. van Breugel, D. Minitti, et al.,
    Astron. Astrophys. 352, L51 (1999).
\bibitem{}
C. De Breuck, F. Bertoldi, C. Carilli, et al., Astron.
   Astrophys. 424, 1 (2004).
\bibitem{}
C. De Breuck, W. van Breugel, H. J. A. Rottgering,
  and G. Miley, Astron. Astrophys., Suppl. Ser. 143,
   303 (2000).

\bibitem{}
 C. De Breuck, W. van Breugel, H. Rottgering, et al.,
     Astron. J. 121, 1241 (2001).
\bibitem{}
 C. De Breuck, W. van Breugel, S. A. Stanford, et al.,
   Astron. J. 123, 637 (2002).
\bibitem{}
 J. N. Douglas, F. N. Bash, and F. A. Bazyan, Astron.
     J. 111, 1945 (1996).
\bibitem{}
   X. Fan, M. A. Strauss, and D. P. Schneider, Astron.
      J. 125, 1649 (2003).
\bibitem{}
W. M. Goss, Yu.N.Parijskij, A.I. Kopylov, et al., Turk.
    J. Phys. 18, 894 (1994).
\bibitem{}
 J. F. R. Gower, P.F.Scott, and D.Wills, Mon. Not. R.
    Astron. Soc. 71, 49 (1967).
\bibitem{}
 A. I. Kopylov, V. M. Goss, Yu. N. Pariiskii, et al.,
    Astron. Zh. 72, 613 (1995) [Astron. Rep. 39, 543
      (1995)].
\bibitem{}
M. Lacy,A.J.Bunker, andS.E.Ridgway,Astron. J.
120, 68 (2000). 
\bibitem{}
 M. Lacy, G. Miley, S. Rawlings, et al., Mon. Not. R.
     Astron. Soc. 271, 504 (1994).
\bibitem{}
  M. Lacy, S. Rawlings, G. J. Hill, et al., Mon. Not. R.
     Astron. Soc. 308, 1096 (1999).
\bibitem{}
 A. Loeb and R. Barcana, Ann. Rev. Astron. Astrophys. 39, 19 (2001).
\bibitem{}
  S. Malhotra and J. E. Rhoads, Astrophys. J. 626, 666
   (2005).
\bibitem{}
   P. J. McCarthy, Ann. Rev. Astron. Astrophys. 31, 639
       (1993).
\bibitem{}
   B. Y. Mills, O. B. Slee, and E. R. Hill, Aust. J. Phys.
       11, 360 (1958).
\bibitem{}
   T. W. B. Muxlow, A. M. S. Richards, S. T. Garrington,
       et al., Mon. Not. R. Astron. Soc. 358, 1159 (2005).
\bibitem{}
Yu. N. Parijskij, N. N. Bursov, N. M. Lipovka, et al.,
	Astron. Astrophys., Suppl. Ser. 87, 1 (1991).
\bibitem{}
 Yu. N. Parijskij, N. N. Bursov, N. M. Lipovka, et al.,
	 Astron. Astrophys., Suppl. Ser. 96, 583 (1992).
\bibitem{}
Yu. N. Parijskij, W. N. Goss, A. I. Kopylov, et al., Bull.
	 Spec.Astrophys.Obs. 40, 5 (1996).
\bibitem{}
Yu. N. Parijskij, W. N. Goss, A. I. Kopylov, et al.,
   Astron. Astrophys. Trans. 18, 437 (1999).
\bibitem{}
R. Pello, D. Shaerer, J. Richard, et al., Astron. Astrophys. 416,
    L35 (2004).
\bibitem{}
S. Rawlings, M. Lacy, K. M. Blundell, et al., Nature
       383, 502 (1996).
\bibitem{}
E. A. Richards, Astrophys. J. 533, 611 (2000).
\bibitem{}
H. J. A. R\"ottgering, M. Lacy, G. R. Miley, et al.,
    Astron. Astrophys., Suppl. Ser. 108, 79 (1994).
\bibitem{}
N. S. Soboleva, O. V. Verkhodanov, et al., Pisma
   Astron. Zh. 26, 723 (2000) [Astron. Lett. 26, 623
   (2000)].

\bibitem{}
A. Songaila, Astron. J. 127, 2598 (2004).
\bibitem{}
H. Spinrad, A. Dey, andJ.R.Graham, Astrophys.J.
     438, L51 (1995).
\bibitem{}
E. R. Stanway, A. J. Bunker, and R. G. McMachon,
     Mon. Not. R. Astron. Soc. 342, 439 (2003).
\bibitem{}
A. G. G. M. Tielens, G. K. Miley, and A. G. Willis,
      Astron. Astrophys., Suppl. Ser. 35, 153 (1979).
\bibitem{}
W. G. M. van Breugel, S. A. Stanford, H. Spinrad,
      et al., Astrophys. J. 502, 614 (1998).
\bibitem{}
B. P. Venemans, H. J. A. Rottgering, R. A. Overzier,
      et al., Astron. Astrophys. 424, L17 (2004).
\bibitem{}
O. V. Verkhodanov, Yu.N.Parijskij, N.S.Soboleva,
      et al., Bull. Spec. Astrophys. Obs. 52, 5 (2001).
\bibitem{}
O. V. Verkhodanov, S. A. Trushkin, H. Andernach,
   and V.N. Chernenkov, Astron.Soc.Pac.Conf. Ser.
   125, 322 (1997).
\bibitem{}
I. Waddington, R. A. Windhorst, S. H. Cohen, et al.,
    Astrophys. J. 526, L77 (1999).
\bibitem{}
F. Walter, C. Carilli, and F. Bertoldi, Astrophys. J. 615,
     L17 (2004).
\bibitem{}
P. J. S. Williams, R. A. Collins, J. L. Caswell, and
     D. J. Holden, Mon. Not. R. Astron. Soc. 139, 289
     (1968).
\end{thebibliography}
\end{document}